\documentclass[aps,prb,twocolumn,floatfix,showpacs]{revtex4}
\usepackage{epsfig}
\begin{document}
\title{Spin transport in inhomogeneous magnetic fields: a
proposal for Stern-Gerlach-like experiments with conduction electrons.
}

\author{Jaroslav Fabian$^{1,2,3}$}

\author{S. Das Sarma$^1$}

\affiliation{$^1$Condensed Matter Theory Center, Department of Physics, University of Maryland at  College
Park, College Park, Maryland 20742-4111, USA\\
$^2$Max-Planck Institute for the Physics of Complex Systems, 01187 Dresden, Germany \\
$^3$ Institute for Theoretical Physics, Karl-Franzens University, 8010 Graz, Austria}

\begin{abstract}
Spin dynamics in spatially inhomogeneous magnetic fields is studied 
within the framework of Boltzmann
theory. Stern-Gerlach-like separation of spin up and spin down
electrons occurs in ballistic and diffusive regimes, before spin
relaxation sets in. Transient dynamics and spectral response
to time-dependent inhomogeneous magnetic fields are investigated,
and possible experimental observations of our findings are discussed.

\end{abstract}

\pacs{72.25.-b, 72.25.Ba, 72.25.Dc, 72.25.Rb}

\maketitle


Spin of mobile carriers (electrons and holes) plays an active central
role in the current spintronics efforts\cite{prinz98,dassarma00}, where
electronic properties are determined, to a great degree,
by the presence of nonequilibrium spin. 
The spin response to temporal and spatial
changes of the magnetic environment determines
various properties of such devices. In the recently
proposed magnetic diode\cite{zutic01a,zutic01b,zutic02}, for example, an inhomogeneous
magnetic environment results from inhomogeneous magnetic
doping. How a magnetic diode adjusts to
the switching of an external magnetic field
and to changes in the applied bias  
depends on the way the mobile
carriers reach equilibrium. Since inhomogeneous
magnetic fields are ubiquitous 
in spintronic devices (mostly due to the presence
of magnetic/nonmagnetic interfaces), it is important to understand
nonequilibrium spin evolution in such fields.
In this paper we investigate in detail the transient 
behavior of conduction electron spins, within a Boltzmann
equation model. A unique feature of the model is that it is
exactly soluble, allowing a detailed study of the transition 
from the ballistic to the diffusive regime. We show that in 
inhomogeneous magnetic fields
a spatial separation between spins, an analogue
of the Stern-Gerlach effect, occurs before
spin relaxation begins, but spin current 
vanishes much sooner, at times of the order of transit times. 

The model we consider is a degenerate electron
gas (a metal or semiconductor) in a magnetic field with
the largest component in the $\hat{\bf z}$ direction, and with a gradient
in that direction. The field has also transverse components (as required
by $\nabla\cdot {\bf B}=0$), which are essential in rendering SG with electron
beams useless\cite{mott29}, 
but which do not hinder an effective spin 
separation of conduction electrons (see below). We show that an 
effective spatial spin separation, along with a flow of 
spin, is possible within ballistic
and diffusive dynamics, demonstrating a Stern-Gerlach-like (SG)
effect with conduction electrons. The formalism we use,  linear 
response theory within the Boltzmann equation, has been applied 
earlier in various forms in transport in general\cite{harrison79,forster75}, 
and more specifically for spin transport in the framework of conduction 
electron spin resonance \cite{walker71,platzman73} and giant 
magnetoresistance\cite{valet93}. Here we apply this
formalism to a special case of spin dynamics in an inhomogeneous
magnetic field, and solve it exactly for specific boundary conditions. 

In the presence of a uniform electric field ${\bf E}$ and 
inhomogeneous magnetic field ${\bf B}$, semiclassical dynamics of 
electrons in (nonmagnetic) 
solids is governed by the Boltzmann equation
\begin{eqnarray}\label{eq:be}
\frac{\partial f_{{\bf k}\lambda}}{\partial t}
+{\bf v}_{\bf k}\cdot\frac{\partial f_{{\bf k}\lambda}}{\partial {\bf r}}
-e{\bf E}_{\lambda}\cdot\frac{\partial f_{{\bf k}\lambda}}
{\partial\hbar
{\bf k}}=-\frac{\delta\bar{f}_{{\bf k}\lambda}}{\tau}-
\frac{\delta f_{{\bf k}\lambda}}{T_{1}},
\end{eqnarray}
where $f_{{\bf k}\lambda} \equiv f_{{\bf k}\lambda}({\bf r},t)$ is
the distribution function 
of electrons with lattice momentum ${\bf k}$ (band index is suppressed) and 
spin $\lambda$ ($1$ or $\uparrow$ for up and $1$ or $\downarrow$ 
for down), at point ${\bf r}$ and time $t$. The notation for
the drift field is simplified as 
${\bf E}_{\lambda}=E\hat{\bf z}+\lambda (\mu_B/e) 
\partial B\hat{\bf z}/\partial z$, 
where $\mu_B$ is the Bohr magneton and the electron $g$-factor is taken to be $2$;
the fields are oriented in the $\hat{\bf z}$ direction. 
Band velocity ${\bf v}_{\bf k}\equiv\partial \varepsilon_{\bf k}/
\partial \hbar {\bf k}$, with $\varepsilon _{\bf k}$ standing for band energy
(we consider systems with inversion symmetry where band energy 
is spin  independent). Two momentum relaxation processes are
distinguished in Eq.~\ref{eq:be}. %
First, spin-conserving momentum scattering
with rate $1/\tau$ ($\tau$ is momentum relaxation 
time), and leading to a quasi-equilibrium distribution 
$\bar{f}_{{\bf k}\lambda}$
($\delta \bar{f}_{{\bf k}\lambda}\equiv
f_{{\bf k}\lambda}-\bar{f}_{{\bf k}\lambda}$), in which
spin up and down electrons have different chemical potentials. 
Second, spin-flip momentum relaxation with rate $1/T_{1}$ 
($T_1$ is spin-relaxation time), and leading to complete
(momentum and spin) equilibrium at the local and instantaneous 
${\bf B}$-field: $f_{{\bf k}\lambda}^0=
f_{0}[\varepsilon_{\bf k}+\lambda\mu_B B(z,t)]$, 
where $f_{0}(\varepsilon)=1/[\exp(\varepsilon-\mu)/k_BT+1]$ 
is the Fermi-Dirac distribution function  
with chemical potential $\mu$, temperature $T$, and Boltzmann constant
$k_B$ ($\delta f_{{\bf k}\lambda}=f_{{\bf k}\lambda}-f_{{\bf k}\lambda}^0$).
In writing the Boltzmann equation as Eq. \ref{eq:be}, we neglect 
the Lorentz force as unimportant, as the largest part of the magnetic field
is oriented along the same direction as the drift velocity itself (see below 
for the reasons why we also neglect the orbital effects of the small 
transverse magnetic
fields).  The dynamics of the transverse spin components ($x$ and $y$)
is also not considered, as it is masked by their fast precession about 
$\bf {B}$.   
Finally, the contribution of the electronic magnetization to the magnetic field
is neglected for our nonmagnetic systems. 
The relaxation time approximation used in Eq. \ref{eq:be}
is good for all practical purposes, but some caution is needed 
especially at low temperatures, as shown in Appendix \ref{appendix:1}.
Generalization 
of Eq. \ref{eq:be} to ${\bf k}$-dependent $\tau$, $T_1$ (which can vary more
wildly over the Fermi surface than $\tau$--see Ref. \onlinecite{fabian98}), 
and $g$-factor is straightforward. 

We search for the solution of Eq. \ref{eq:be} in the form 
$f_{{\bf k}\lambda}=f^0_{{\bf k}\lambda}-(\partial f_{0}/\partial
\varepsilon_{\bf k})\phi_{{\bf k}\lambda}$, and write the
quasi-equilibrium distribution function as $\bar{f}_{{\bf k}\lambda}=
f_{{\bf k}\lambda}^0-(\partial f_{0}/\partial \varepsilon_{\bf k})
\mu_{\lambda}$, where the nonequilibrium chemical potential
$\mu_{\lambda}=\langle \phi_{{\bf k}\lambda} \rangle$
is obtained self-consistently from the integral relation for 
spin conservation,
$\sum_{\bf k}\delta\bar{f}_{{\bf k}\lambda}=0$. The 
angular brackets introduce the Fermi-surface averaging:
$\langle ... \rangle \equiv \sum_{\bf k} ... 
\partial f_0/\partial \varepsilon_{\bf k}/\sum_{\bf k} \partial f_0/\partial 
\varepsilon_{\bf k}$. 
After linearizing Eq. \ref{eq:be} in terms of $\phi$ and $E_{\lambda}$,
we obtain
\begin{eqnarray}\label{eq:bel}
-\lambda \mu_B\frac{\partial B}{\partial t}
+\frac{\partial \phi_{{\bf k}\lambda}}{\partial t}
+v_{z{\bf k}}\frac{\partial \phi_{{\bf k}\lambda}}{\partial z} +
eEv_{z{\bf k}}=I(\phi_{{\bf k}\lambda}),
\end{eqnarray}
with the collision integral 
\begin{eqnarray}\label{eq:col}
I(\phi_{{\bf k}\lambda})=
-\frac{\phi_{{\bf k}\lambda}-\langle\phi_{{\bf k}\lambda}\rangle}
{\tau}-\frac{\phi_{{\bf k}\lambda}}{T_{1}}.
\end{eqnarray}
Particle number conservation requires
that $\langle \phi_{{\bf k}\uparrow} +\phi_{{\bf k}\downarrow} \rangle$
vanishes. The total spin density is 
\begin{equation}\label{eq:spin}
S=-g_F \mu_B B + 
(1/2) g_F \langle 
\phi_{{\bf k}\uparrow}-\phi_{{\bf k}\downarrow} \rangle,
\end{equation}
where $g_F=-2\sum_{\bf k}\partial f_0/\partial \varepsilon_{\bf k}$
is the density of states, per unit volume, at the Fermi level. The first
term on the RHS of Eq. \ref{eq:spin} is
the equilibrium spin value, yielding the electron gas paramagnetic susceptibility of
$\mu_B^2 g_F$, while the second part, $\delta S$, represents 
the nonequilibrium contribution to spin density. The spin current density
is 
\begin{equation}
J_s=(1/2)g_F\langle v_{z\bf k}
(\phi_{{\bf k}\uparrow}-\phi_{{\bf k}\downarrow}) \rangle, 
\end{equation}
and is connected to $S$ via the continuity equation derived from Eq. \ref{eq:bel}, 
\begin{equation}
\partial S/\partial t +\partial J_s/\partial z=-\delta S/T_1,
\end{equation}
which, together with the linear response equation 
(see Ref. \cite{johnson87} for a systematic treatment of 
linear spin transport), 
\begin{equation}\label{eq:Js}
J_s=-D\partial \delta S/\partial z, 
\end{equation}
where $D=\langle v^2_{z\bf k} \rangle \tau$ is the electron diffusivity
constant, gives the
diffusion formula for investigating diffusive spin transport,
\begin{eqnarray}\label{eq:dif}
\frac{\partial \delta S}{\partial t} -D\frac{\partial^2 \delta S}
{\partial z^2}
=-\frac{\delta S}{T_1}-\frac{\partial S_0}{\partial t}.
\end{eqnarray}

We first study transient phenomena that describe  
evolution of $S$ and $J_s$ towards equilibrium. 
Consider an unpolarized sample (whose 
band structure is assumed isotropic,
$v_{\bf k}=\hbar {\bf k}/m$, where $m$ is electron band mass),
stretching from $0$ to $L$ along the $z$-axis, 
with no charge current 
($E=0$). At $t=0$, magnetic field $B(z)=B_0+B_1z$ is applied.
Our goal is to find, by solving Eq. \ref{eq:bel}, 
$\langle \phi \rangle\equiv \langle \phi_{\uparrow} \rangle=-
\langle \phi_{\downarrow} \rangle$ [so that $S=-g_F\mu_B B+
g_F\langle \phi  \rangle]$, subject
to the initial condition $\langle \phi(z,0)\rangle =\mu_B [B(z)]$, where 
$[B(z)]=B_0+B_1[z]$ is the even periodic extension of $B(z)$ from
interval $(0,L)$ to the whole $z$-axis; thus formulated
initial condition guarantees that spin current vanishes at the 
boundary--the assumption well
justified in nonmagnetic interfaces with negligible spin-flip
scattering. The spin profile can be written as
$S=S_h+S_{in}$, where
the homogeneous $S_h$ and inhomogeneous $S_{in}$ spin components are
\begin{eqnarray}
S_{h}(t)&=&-g_F\mu_B(B_0+\frac{1}{2}B_1L)\left [1-K(0,t)\right ], \\
S_{in}(z,t)&=&g_F\mu_B B_1 4L 
\sum_{n\ge 0}\frac{\cos(q_nz)}{q_n^2L^2}\left [1-K(q_n,t)\right ],
\end{eqnarray}
with $q_n\equiv (2n+1)\pi/L$. The sum comes from 
the Fourier expansion of $[B(z)]$. Kernel 
$K(q,t)$ describes the time evolution of the Fourier $q$-components of 
the nonequilibrium spin:
\begin{equation} 
K(q,t)\equiv\delta S(q,t)/\delta S(q,0)= \langle \phi(q,t) \rangle/\langle 
\phi(q,0) \rangle.
\end{equation}
Having the spin, the spin current can be calculated from the 
continuity equation as 
\begin{equation}
J_s(z,t)=-\int_0^zdz'(\partial /\partial t+1/T_1)\delta S(z',t).
\end{equation}
Equation \ref{eq:bel} gives also an exact {\it quasilocal}
relationship between spin current and spin, valid at all times, and expressed
in terms of the Fourier coefficients as
\begin{eqnarray} \label{eq:ql}
J_s(q,t)=iv_F\delta S(q,0)R_1(q,t)+
iv_F\frac{1}{\tau}\left [\delta S*R_1\right ](q,t).
\end{eqnarray}
Here $v_F$ is the Fermi velocity, 
\begin{equation}
R_1(q,t)=\exp(-t/\tau_m)\frac{d[\sin(x)/x]}{dx},
\end{equation}
 with $x=qv_Ft$ and 
$1/\tau_m=1/\tau+1/T_1$ the total momentum scattering rate, and
the star symbol denotes temporal convolution:
\begin{equation}
[f_1*f_2](q,t)\equiv \int_0^tdt'f_1(q,t-t')f_2(q,t').
\end{equation}
In real space Eq. \ref{eq:ql} expresses $J_s$ in terms of derivatives 
(in principle of all odd orders--that is why the term quasilocal) of
$\delta S$.  In the diffusive regime, at
$t\gg \tau$, the memory of the initial condition is lost, and
Eq. \ref{eq:ql} reduces to Eq. \ref{eq:Js}.  An exact generalization of
the diffusion equation \ref{eq:dif} is obtained by substituting 
$J_s$ from Eq. \ref{eq:ql} to the continuity
equation. The result is
\begin{eqnarray}\label{eq:ie}
K(q,t)=R_0(q,t)+ \frac{1}{\tau}\left [K*R_0\right](q,t),
\end{eqnarray}
where now 
\begin{equation}
R_0(q,t)=\exp(-t/\tau_m)\sin(qv_Ft)/(qv_Ft).
\end{equation}
At $t>\tau$, Eq. \ref{eq:ie}  is equivalent to Eq. \ref{eq:dif}.

Equation \ref{eq:bel} can be solved exactly with the help of Laplace
transform. The solution is provided in Appendix \ref{appendix:2}. The
result is 
\begin{eqnarray}\label{eq:sol}
K(q,t)=
(qv_F\tau)^2e^{-t/\tau_m}\left |\frac{\exp\left[qv_Ft\cot\left(qv_F\tau\right)\right]}
{\sin\left(qv_F\tau\right)^2}\right |^{\rm sing}_{qv_F\tau},
\end{eqnarray}
where the vertical bars denote the singular (principal) part 
of the Laurent series in terms of $qv_F\tau$, of the expression inside. 
An alternative formulation for the Kernel is  
\begin{eqnarray} \label{eq:sol1}
K(q,t)= -\left(\tau/t\right)e^{-t/\tau_m}\sum_{n=-\infty}^{-1}nF_n(qv_Ft)(qv_F\tau)^n,
\end{eqnarray}
where functions $F_n(x)$ are described in Appendix \ref{appendix:2}.

Let us consider the limiting behavior of $K(q,t)$ for the
ballistic and diffusive regimes. For ballistic transport,
$t \ll \tau$, the evolution kernel, Eq. \ref{eq:sol},
reduces to
\begin{eqnarray} \label{eq:spinb}
K_{\rm ball}(q,t)=\frac{\sin(qv_Ft)}{qv_Ft}.
\end{eqnarray}
This is the solution of Eq. \ref{eq:bel} in the absence of scattering.
A finite $S(z,t)$ in the ballistic case is solely due to SG effect
of semiclassical separation of spins. At the left boundary,
\begin{equation}
S_{\rm ball}(0)=-g_F\mu_B B_1L (v_F t/2L),
\end{equation}
in the middle $S_{\rm ball}(L/2)=0$, and at
the right boundary $S_{\rm ball}(L)=-S_{\rm ball}(0)$. The spin separation
$S(L)-S(0)$ grows linearly with time, reaching its maximum of about
$-g_F\mu_B B_1 \ell$ at $t=\tau$. For the diffusive transport,
$t\gg \tau$, Eq. \ref{eq:bel} gives $K\approx K_{\rm diff}$, where
\begin{eqnarray}\label{eq:spind}
K_{\rm diff}(q,t)=\exp(-q^2Dt-t/T_1),
\end{eqnarray}
which is also a solution of Eq. \ref{eq:dif}.
A new time scale, $t_T=L^2/D\pi^2$, appears, for the transit time of a
diffusing electron crossing the sample ($\pi^2 t_T$ is
called the  Thouless time). We consider $L$
smaller than the spin diffusion length
(which can be as large as a millimeter\cite{silsbee80}), so that
$t_T<T_1$.
For $t<t_T$ the spin density grows diffusively,
\begin{equation}
S_{\rm diff}(0)= g_F\mu_B B_1 L 2(Dt/L^2\pi)^{0.5},
\end{equation}
$S_{\rm diff}(L)=-S_{\rm diff}(0)$, while $S_{\rm diff}(L/2)=0$; a large
spin current flows in the middle of the
sample. While the spin current vanishes at greater times, $t>t_T$,
(when drift is being balanced by diffusion), an effective spin separation
remains almost stationary until $t=T_1$,
when spin relaxation establishes equilibrium.
Note that the homogeneous component of spin, $S_h$, evolves towards
equilibrium with spin-flip processes only, since $K(0,t)=\exp(-t/T_1)$
at all times.

\begin{figure}
\centerline{\psfig{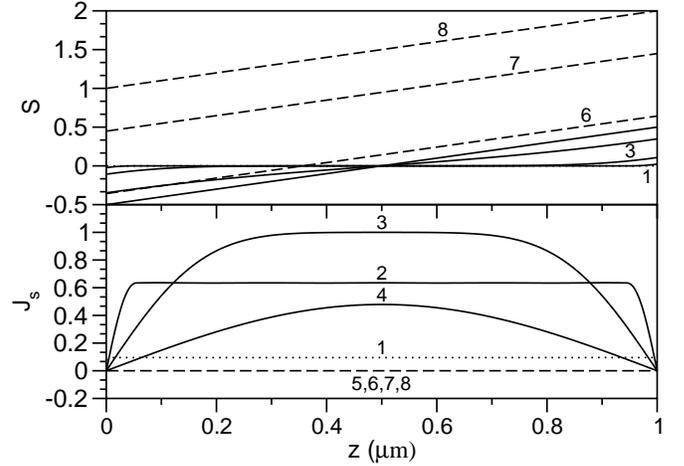}}
\caption{Calculated normalized spin density $\tilde{S}$
(top) and spin current $\tilde{J}_s$ (bottom) for a model
sample defined in the text. The curves represent profiles
at times $t=10^{-14}$, ...,$10^{-7}$ s (increasing by a decade),
and are denoted by a corresponding number 1,...,8 (except
for a few cases at the top, where the trend is clear). The
dotted lines represent the initial ballistic transport at
$t=10^{-14}$ s, while the long-dashed lines are for the
longest times.
}
\label{fig:1}
\end{figure}

\begin{figure}
\centerline{\psfig{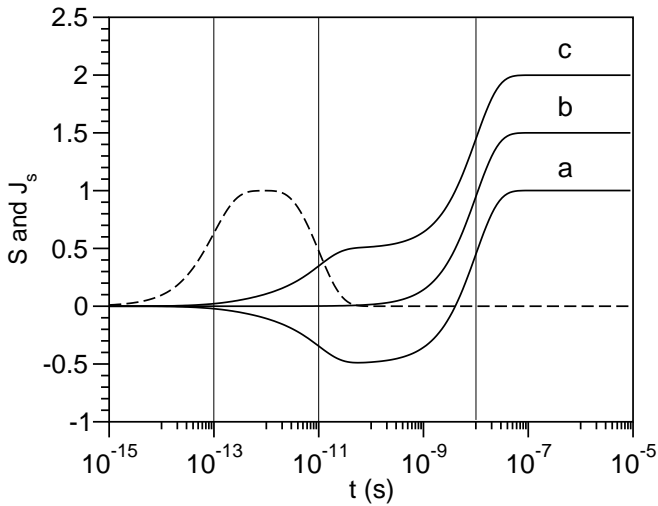}}
\caption{Calculated time evolution of normalized spin density
$\tilde{S}$ at $z=0$ (a), $z=L/2$ (b) and $z=L$ (c). The dashed
line is the time evolution of normalized spin current $\tilde{J}_s$
at $z=L/2$. The three vertical lines separate the ballistic
($t<\tau=10^{-13}$s), diffusive ($\tau < t < t_T\approx 10^{-11}$ s),
quasiequilibrium
($t_T < t < T_1=10^{-8}$ s), and spin relaxational ($t>T_1$) regimes.
}
\label{fig:2}
\end{figure}

An example of a transient evolution of spin and spin current is shown in
Figs. \ref{fig:1} and \ref{fig:2}. We take a model sample of size $L=1\mu$m, 
with realistic electronic parameters $\tau=0.1$ ps, $D=0.01$ m$^2$s$^{-1}$, 
$v_F=(3D/\tau)^{0.5}\approx 5.5\times 10^{5}$ ms$^{-1}$, $t_T=L^2/D\pi^2
\approx 10$ ps, and $T_1=10$ ns. Magnetic field is normalized to $B_0=B_1L$. 
We evaluate our exact solution, Eq. \ref{eq:sol}, numerically  
to obtain spin, and then calculate the spin current from the continuity equation.
The
physics that emerges from our calculation, and which can
be seen on the model example in Figs. \ref{fig:1} and \ref{fig:2}, is
the following. There are four time scales to consider 
(Fig. \ref{fig:2}). (i) In the ballistic
regime ($t<\tau$), electron spin density at the edges begins 
to grow as $\sim t$,
as electrons with one spin direction after bouncing off the boundary 
decelerate and stay close, while the electrons with the opposite spin
accelerate in the other direction. Spin current, which is
always largest in the middle of the sample, rapidly increases to reach
its maximum value at $t=\tau$ (see Fig. \ref{fig:2}). Note that
positive $\tilde{J}_s$ means negative $J_s$, and largely a drift spin
flow (spin diffusion acts in the other direction). 
(ii) The diffusive regime ($\tau < t < t_T$) is 
characterized by a further build-up of spin density at the edges of the 
sample (by diffusion) at the rate $\sim t^{0.5}$. This is accompanied by 
a decay of spin current, as the initial drift is now being balanced by 
diffusion. We call this diffusive SG effect. (iii) The quasiequilibrium
regime ($t_T < t < T_1$), where momenta are in equilibrium characterized
by a finite difference in the chemical potentials of spin up and down
electrons, $\delta S$ is spatially uniform so that no spin currents flow, 
and spin densities remain almost constant in time. (iv) Finally, in
the spin relaxation regime ($t>T_1$), the uniform nonequilibrium spin density 
vanishes and complete equilibrium is established. 

\begin{figure}
\centerline{\psfig{file=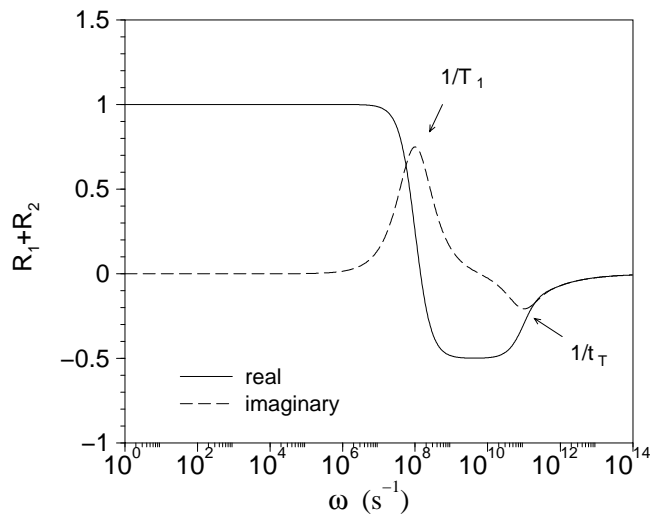,width=1\linewidth,angle=0}}
\caption{Calculated dimensionless spectral response $R_1(\omega)+R_2
(z,\omega)$ at $z=0$ for the model sample.
The two shoulders
in the real part of the response (and the corresponding peaks in the
imaginary part) correspond to spin relaxation ($\omega=1/T_1
=10^8$ s$^{-1}$) and diffusion ($\omega=1/t_T\approx 10^{11}$
s$^{-1}$).
}
\label{fig:3}
\end{figure}

We now ask the question of how the electron spin system responds to a
time-varying inhomogeneous magnetic field. Suppose 
$B(z,t)=(B_0+B_1z)\exp(-i\omega t)$. We show the results for diffusive 
dynamics, and solve Eq. \ref{eq:dif} with $-\partial
S_0/\partial t$ as the source term. Linear-response theory for magnetic
susceptibility for diffusive transport is well known \cite{platzman73};
here we illustrate it for the specific boundary conditions of our
model of spin separation, where various time scales discussed above will
be manifest on the frequency domain.  In response to the oscillating 
field, spin density changes as $\tilde{S}=
[R_1(\omega)+R_2(z,\omega)]\exp(-i\omega t)$, where 
\begin{eqnarray}\label{eq:R1}
R_1&=&\left(\frac{B_0}{B_1L}+\frac{1}{2}\right ) 
\frac{1}{-i\omega T_1+1}, \\ \label{eq:R2}
R_2&=&-4\sum_{n\ge0}\frac{\cos(q_nz)}{q_n^2L^2}
\frac{Dq_n^2T_1+1}{-i\omega T_1+Dq_n^2T_1+1}.
\end{eqnarray}
Spin relaxation is primarily taken over by $R_1$, which measures the
response of the uniform ($q=0$) components of the spin density.
On the other hand, $R_2$ collects the terms responsible for 
diffusion, as diffusion modes ($q > 0$) are the first ones
to achieve equilibrium ($t_T \ll T_1$). The total response, calculated
for our model system, is displayed in Fig. \ref{fig:3}. At small 
frequencies spins can adiabatically (in equilibrium) follow the
local and instantaneous $B(z,t)$: at low $\omega$, 
$\tilde{S}\approx(B_0/B_1L+z/L)\exp(-i\omega t)$. 
At greater frequencies, first the
spin-relaxation peak (in the $\Im{R_1}$) and shoulder (in $\Re{R_1}$) 
appear, while at $\omega\approx 1/t_T$, a second peak and shoulder
(now due to $R_2$) appear, as the time scale of the diffusive regime
is reached. The second peak signals the SG effects, where dissipation
is due to drift spin currents; $\tilde{S}$
reaches negative values of $-1/2$, showing spin separation
(compare with Fig. \ref{fig:2}).
Spectral response of the spin current is 
$\tilde{J_s}(t)=(1-\partial R_2/\partial z)\exp(-i\omega t)$, 
and shows a structure only around 
$\omega \approx 1/t_T$, as spin current relaxes during the 
transit time (and {\it not} on $T_1$ scale). 

Finally, we discuss some issues related to a possible experimental observation
of our findings. 
What we call a SG-like effect is an effective spin (not particle) separation
of electrons in metals and semiconductors. 
Let us summarize the time scales involved. Ballistic transport lasts for 
femtoseconds up to a picosecond, diffusive transit across a micron sample
can take from a picosecond to a nanosecond, and spin relaxation times
can be between a fraction of a nanosecond to a microsecond\cite{fabian99}. 
Ordinary SG fails to work with electrons because transverse magnetic fields 
(say, $B_y=-B_1y$) give rise to the Lorentz force which makes, say, moving to 
the left spin up electrons turn around and move to the right,
smearing out spin separation\cite{mott29} (see, however, Ref. \cite{batelaan97}). 
In our case the time scale of the Larmor precession, $t_L=m/eB$, is large
enough to be neglected. Indeed, for such a large $B_1$ as 10 T/cm, 
the transverse
field would be of order 10 Gauss for a micron sample, turning an electron
around in $t_L\approx 5$ ns, long after spin separation sets in.  
In addition, orbital effects are inhibited due to momentum scattering. One
can still go to a one- or two-dimensional sample to study SG with
ballistically propagating electrons\cite{wrobel01}, 
if Larmor precession is faster than 
momentum relaxation. 

A SG-like spin separation should be observable in both metals and
semiconductors. (We are not aware of any experimental method of
measuring directly the spin current, in our case the diffusive
regime, although theoretical proposals exist--see Ref. \cite{dyakonov71}).
One way of measuring a nonequilibrium spin in metals is the 
Silsbee-Johnson method of spin-charge coupling \cite{silsbee80}.
One can either switch an external inhomogeneous magnetic field, {\it or} inject
nonequilibrium spin into a metal in a static field, to measure
the time evolution of the spin. This can be accomplished, for example,
by placing a ferromagnetic electrode on the top of a sample's edge,
and measure the voltage across the interface, which is proportional
to the nonequilibrium spin \cite{silsbee80}
(the spin in the ferromagnetic
electrode can be considered to be in equilibrium, since
it relaxes much faster than in the nonmagnetic sample).
The voltage would be 
present even when the spin current vanishes (that is, in the
quasiequilibrium regime), and thus can be monitored with the
sub-T$_1$ (not $t_T$) resolution.  Gradient $B_1$ can 
create spin at the sample edges of about $\mu_B B_1 L/E_F$ 
spins per electron ($E_F$ is the Fermi energy), which, 
for typical values of, say, $B_1\approx 1$ T/cm, $L=10$ $\mu$m, 
and $E_F\approx 10$ meV gives about 1 spin per 10$^8$ electrons
(note that $L$ must be smaller than the spin diffusion length to observe
the separation). For comparison, in the Johnson-Silsbee spin 
injection experiment 1 spin per 10$^{11}$ electrons was
detected\cite{silsbee80}.

In semiconductors like GaAs, the traditional tool to observe spin
polarization of the carriers has been photoluminescence 
polarization detection\cite{orientation84}. A degenerate
semiconductor with $E_F\approx 1$ meV (and otherwise
the same conditions as above) would be polarized to 
about 0.01\% (for more sensitivity a greater $B_1$ or $L$, 
or a material with a larger $g$-factor could be used), 
emitting light with circular
polarization of the same order. If the sample is $n$-doped,
for example, and the edges form the interfaces with a $p$-doped
material, the spin polarization of the light emitted
at the edges would be opposite in the quasiequilibrium
regime, demonstrating the SG-separation. Pico-to-micro-second 
resolved pump and probe photoluminescence measurements 
in an inhomogeneous but {\it static} magnetic field could follow
the evolution from ballistic regime to full equilibrium
of a semiconductor spin system, yielding information not
only about spin, but also about charge transport, as seen
from our calculation. In addition to the optical technique
and the Johnson-Silsbee method, one could also in principle 
observe our predicted effect using the magnetic resonance
force microscopy.

In summary we have studied transient spin dynamics of itinerant
electrons in metals and degenerate semiconductors placed 
in an inhomogeneous magnetic field. In particular, we have 
solved exactly a spin dynamics model based on the Boltzmann equation,
and demonstrated that the spin evolution proceeds 
through four distinct modes: ballistic, diffusive, 
quasiequilibrium, and 
equilibrium. An effective spin separation is possible in the 
quasiequilibrium regime, where the spin current vanishes and 
the spin is in equilibrium only with the inhomogeneous component of 
the magnetic field.

We thank Igor \v{Z}uti\'{c} and Xuedong Hu for useful discussions.
This work was supported by the U.S. ONR and DARPA.

\appendix 

\section{\label{appendix:1}}

To demonstrate the effect of spin-flip scattering on charge and spin transport, 
consider electrons in a simple metal,
scattering elastically off impurities at the rate
$W_{{\bf k}\lambda,{\bf k}'\lambda'}$. Spin-flip events are characterized
by $W_{{\bf k}\uparrow,{\bf k}'\downarrow}$ and result mainly from the spin-orbit
interaction. The collision integral is 
\begin{eqnarray}\label{eq:fbe} 
\sum_{{\bf k}'\lambda'}\left[W_{{\bf k}'\lambda',{\bf k}\lambda} f_{{\bf k}'\lambda'}
(1-f_{{\bf k}\lambda})-W_{{\bf k}\lambda,{\bf k}'\lambda'}
f_{{\bf k}\lambda}(1-f_{{\bf k}'\lambda'})\right].
\end{eqnarray}
The first term describes scattering from
${\bf k}'\lambda'$ to ${\bf k}\lambda$, which increases $f_{{\bf k}\lambda}$, while
the second term represents reversed processes, which decrease $f_{{\bf k}\lambda}$.
 Factors like $f_{{\bf k}\lambda}
(1-f_{{\bf k}'\lambda'})$ ensure that the initial state is occupied and final
state empty, in accord with the Pauli principle.

Let the system is driven off equilibrium by an applied electric field $\bf E$ and 
an inhomogeneous chemical potential $\lambda \delta\mu$. The latter does not disturb the 
equilibrium electronic density, but maintains an inhomogeneous nonequilbrium spin 
polarization ($\lambda \delta \mu$ is essentially the driving term for spin diffusion
caused, for example, by spin injection). We seek the
solution to Eq. \ref{eq:be} with the RHS Eq. \ref{eq:fbe} in the form
$ f_{{\bf k}\lambda}=f_{\bf k\lambda}^0- (\partial f_0/\partial \epsilon_{\bf k})
\phi_{{\bf k}\lambda}$, where now $f_{\bf k\lambda}^0=f_0(\epsilon_{\bf k}-
\lambda \delta \mu)$. After linearization the Boltzmann equation becomes 
\begin{eqnarray}\label{eq:lbe}
\left(e{\bf E}+\lambda {\bf \nabla} {\delta\mu}\right)\cdot {\bf v}_{\bf k}
= \sum_{{\bf k}'\lambda'}W_{{\bf k}\lambda,{\bf k}'\lambda'}\left(\phi_{{\bf k}
\lambda}
-\phi_{{\bf k}'\lambda'}\right),
\end{eqnarray}
where we left out the Lorentz force as unimportant for the present discussion.
Equation \ref{eq:lbe} has the formal solution
\begin{equation}
\phi_{{\bf k}\lambda}=-e{\bf E}\cdot {\bf a}_{\bf k} - \lambda 
{\bf \nabla}\delta \mu\cdot {\bf b}_{\bf k},
\end{equation}
with ${\bf a}_{\bf k}$ and ${\bf b}_{\bf k}$ satisfying the integral equations:
\begin{eqnarray}\label{eq:a}
{\bf v}_{\bf k}=\sum_{{\bf k}'} \left[ W_{{\bf k}\uparrow, {\bf k}'\uparrow}
({\bf a}_{{\bf k}'}-{\bf a}_{{\bf k}}) + W_{{\bf k}\uparrow, {\bf k}'\downarrow}
({\bf a}_{{\bf k}'}-{\bf a}_{{\bf k}})\right],
\end{eqnarray}
and
\begin{eqnarray}\label{eq:b}
{\bf v}_{\bf k}=\sum_{{\bf k}'} \left[W_{{\bf k}\uparrow, {\bf k}'\uparrow}
({\bf b}_{{\bf k}'}-{\bf b}_{{\bf k}}) - W_{{\bf k}\uparrow, {\bf k}'\downarrow}
({\bf b}_{{\bf k}'}+{\bf b}_{{\bf k}})\right].
\end{eqnarray}
Vectors ${\bf a}_{\bf k}$ and ${\bf b}_{\bf k}$ have magnitudes of order $\ell$,
and their knowledge allows to calculate the tensors of charge and spin
conductivities: $ \sigma=2e^2\sum_{\bf k} (-\partial f_{\bf k}^0/\partial
\epsilon_{\bf k}){\bf v}_{\bf k} {\bf a}_{\bf k}$ and
$\sigma_S=2e^2\sum_{\bf k} (-\partial f_{\bf k}^0/\partial
\epsilon_{\bf k}){\bf v}_{\bf k} {\bf b}_{\bf k}$. The spin conductivity
$\sigma_S$ is related to the spin diffusivity $D_S$ as $\sigma_S=e^2 g_F D_S$, since the 
nonequilibrium spin is $\delta S=g_F\delta \mu$. Here we use spin 
conductivity instead of the more usual spin diffusivity only to 
stress the contrast with charge conductivity.

If there is no spin-flip scattering, ${\bf a}_{\bf k}={\bf b}_{\bf k}$,
the effective mean free paths are the same for both currents, and
the conductivities are equal: $\sigma_S=\sigma$.
Spin-flip scattering, however, implies ${\bf a}_{\bf k}\ne
{\bf b}_{\bf k}$, and so
it plays different roles in charge and spin transport.
Assume, for a moment, that scattering is isotropic, and energy surfaces
spherical. Then Eqs. \ref{eq:a} and \ref{eq:b} can be solved exactly \cite{ziman60}
by introducing transport relaxation times
$\tau$ and $\tau_S$: ${\bf a}_{\bf k}=-\tau{\bf v}_{\bf k}$ and ${\bf b}_{\bf k}=
-\tau_S{\bf v}_{\bf k}$, and, with $\theta$ being the angle between
${\bf v}_{\bf k}$ and ${{\bf v}_{\bf k}'}$,
\begin{eqnarray}\label{eq:tau}
\frac{1}{\tau}=\sum_{{\bf k}'} \left[W_{{\bf k}\uparrow,{\bf k}'\uparrow}
                           (1-\cos\theta)+
                          W_{{\bf k}\uparrow,{\bf k}'\downarrow}
                          (1-\cos\theta)\right],
\end{eqnarray}
and
\begin{eqnarray}\label{eq:taus}
\frac{1}{\tau_S}=\sum_{{\bf k}'} \left[W_{{\bf k}\uparrow,{\bf k}'\uparrow}
(1-\cos\theta)+ W_{{\bf k}\uparrow,{\bf k}'\downarrow}
                          (1+\cos\theta)\right].
\end{eqnarray}

In charge transport, spin-conserving and spin-flip processes contribute
in the same way: they are weighted by the well known $1-\cos\theta$, which suppresses
contributions from small-angle scattering as ineffective in degrading
charge current. Spin transport is a different story. Here
spin-flip processes come with $1+\cos\theta$, and backscattering
($\theta\approx \pi$) is the least effective in degrading spin current, while
small-angle events contribute most. There is illuminating physics behind this:
Spin up and down electrons move antiparallel to each other, so if any
spin flip is accompanied by the velocity reversal,
 the current does not change. But if the velocity
stays the same, the effect is maximal, as if the
electron spin does not flip, but the velocity reverses.
Equations \ref{eq:tau} and \ref{eq:taus} are not valid in more
general cases of anisotropic bands and inelastic anisotropic scattering,
but, with some caution, they are still a useful approximation
(justified by variational analysis\cite{ziman60}) in the interesting case
of electrons scattering by thermal fluctuations (lattice or spin) at low $T$.
Such fluctuations will allow only small-angle
scattering, suppressing contributions from  spin-conserving, but not from
spin-flip processes.  Spin-flip processes are thus much more important for spin
transport than for charge transport. For example, the contribution of the
phonon-induced
spin-flip scattering to $1/\tau$ falls as $T^7$, while to $1/\tau_S$ only as $T^5$
(this follows from Yafet's theory \cite{yafet63} as also confirmed by 
a numerical calculation  \cite{fabian99}). The contribution
from the spin-conserving electron-phonon interaction falls
as $T^5$ (the Bloch-Gr\"{u}neisen law), for both charge and spin currents.

\section{\label{appendix:2}}

\begin{figure}
\centerline{\psfig{file=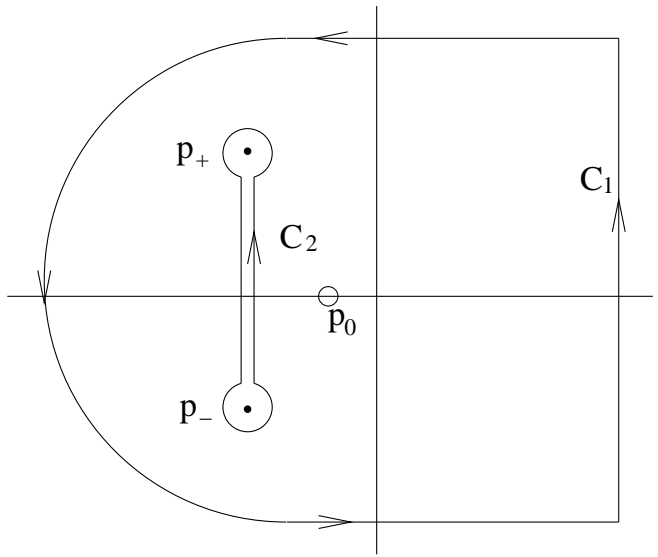,width=1\linewidth,angle=0}}
\caption{Integration contour for $K(q,t)$ of Eq. \ref{eq:lap} with the horizontal 
and vertical axes representing the real and imaginary part of $p$, respectively. 
The integral in Eq. \ref{eq:lap}, which runs along $C_1$, is
the same as the integral along the path $C_2$ cutting out the branch
line from $p_-$ to $p_+$, plus the residue at $p_0$. 
}
\label{fig:4}
\end{figure}

The solution of Eq.  \ref{eq:bel} (or, equivalently, Eq. \ref{eq:ie}) can be 
written with the help of Laplace transform as the integral in the complex
plane:
\begin{eqnarray} \label{eq:lap}
K(q,t)=\int_{-\infty+i\sigma}^{\infty+i\sigma}
\frac{dp}{2\pi i}\frac{e^{pt}\langle\left(p+iqv+1/\tau_m\right)^{-1}\rangle}
{1-\langle\left(p+iqv+1/\tau_m\right)^{-1}\rangle/\tau},
\end{eqnarray}
where $\sigma>0$. For a degenerate system considered in the text,
\begin{eqnarray}
\langle\left(p+iqv+1/\tau_m\right)^{-1}\rangle=\frac{1}{2iqv_F}
\ln\left[\frac{p+iqv_F+1/\tau_m}{p-iqv_F+1/\tau_m}\right ].
\end{eqnarray} 
The integral can be evaluated by a suitable contour deformation in the complex plane,
as indicated in Fig. \ref{fig:4}. The original integral in Eq. \ref{eq:lap}, which goes
along $C_1$ is the same as the integral over $C_2$ plus the residue at $p_0$. The 
path $C_2$ cuts away the branch line extending from $p_-=-1/\tau_m-iqv_F$ to 
$p_+=-1/\tau_m+iqv_F$, from the complex plane. The residue is evaluated for the
pole at $p_0=-1/\tau_m+qv_F\cot(qv_F\tau)$ present for $qv_F\tau\le \pi/2$ (defining
the Riemann sheet for $\arctan$ to go from $-\pi/2$ to $\pi/2$). 

The result of the contour integration can be formally written as 
\begin{eqnarray}\label{eq:k}
K(q,t)= 
(qv_F\tau)^2e^{-t/\tau_m}\left |\frac{\exp\left[qv_Ft\cot\left(qv_F\tau\right)\right]}
{\sin^2\left(qv_F\tau\right)}\right |^{\rm sing}_{qv_F\tau}, 
\end{eqnarray}
where the vertical bars denote the singular (principal) part
of the Laurent series in terms of $qv_F\tau$, of the expression inside.
An alternative formulation is 
\begin{eqnarray} \label{eq:kfn}
K(q,t)= -\left(\tau/t\right)e^{-t/\tau_m}\sum_{n=-\infty}^{-1}nF_n(qv_Ft)(qv_F\tau)^n,
\end{eqnarray}
where functions $F_n(x)$ are the coefficients of the Laurent
series:
\begin{equation}
\exp[x\cot(y)]=\sum_{n=-\infty}^{\infty}F_n(x)y^n,
\end{equation}
satisfying the recursion relation
\begin{eqnarray}\label{eq:rec}
F_n(x)+F''_n(x) =-\frac{n+1}{x}F_{n+1}(x),
\end{eqnarray}
with the boundary conditions $F_n(0)=0$ and $F'_n(0)=\delta_{n,-1}$ for $n \le -1$.
In principle, all the functions $F_n(x)$, $n=-2,..,-\infty$ can thus be generated
from 
\begin{equation}
F_{-1}(x)=\sin(x),
\end{equation} which is readily obtained from Eq. \ref{eq:rec}.

The limiting case for ballistic transport can be obtained from 
Eq. \ref{eq:kfn} by letting $qv_F\tau$ (and thus also $qv_FT_1$) going to infinity. The result 
is 
\begin{equation}
K(q,t) \approx K_{\rm ball}(q,t)=\frac{F_{-1}}{qv_F t}.
\end{equation}
On the other hand, the diffusive limit can be obtained by letting $qv_F\tau$ to 
zero, in which case the vertical bar in Eq. \ref{eq:k} can be removed (the singular
part equals the whole part since the regular part vanishes). By expanding
$\cot(qv_F\tau)\approx 1/(qv_F\tau)-(qv_F\tau)/3$ and denoting as $D\equiv 
v_F^2\tau/3$, one obtains
\begin{equation}
K(q,t) \approx K_{\rm diff}(q,t)= \exp(-t/T_1-q^2Dt).
\end{equation}

\end{document}